\documentclass[11pt]{article}
\usepackage{epsfig}
\usepackage{graphicx}
\usepackage{amsmath}

\newcommand{\BABARPubYear}    {01}

\newcommand{\BABARProcNumber} {68}
\newcommand{\SLACPubNumber} {9031}

\input pubboard/babarsym

\long\def\inst#1{\par\nobreak\kern 4pt\nobreak
    {\it #1}\par\vskip 10pt plus 3pt minus 3pt}

\begin{document}
{\pagestyle{empty}

\begin{flushright}
SLAC-PUB-\SLACPubNumber \\
\babar-PROC-\BABARPubYear/\BABARProcNumber \\
October, 2001 \\
\end{flushright}

\par\vskip 4cm

\begin{center}
\Large \bf Observation of CP violation in the $B^0$ system
\end{center}
\bigskip

\begin{center}
\large
    Christos Touramanis\\
    Department of Physics\\
    Oliver Lodge Laboratory, University of Liverpool\\
    L69 7ZE, Liverpool, U.K.\\
(for the \lbabar\ Collaboration)
\end{center}
\bigskip \bigskip

\begin{center}
\large \bf Abstract
\end{center}
The \babar\ detector, operating at energies near the $\FourS$
resonance at the \pep2\ asymmetric \BF\ at SLAC, has collected a
sample of 32 million $B\Bbar$ pairs by May 2001. A study of
time-dependent \CP-violating asymmetries in events where one
neutral $B$ meson is fully reconstructed in a final state
containing charmonium has resulted in the measurement $\stwob=0.59
\pm 0.14\ {\rm (stat)} \pm 0.05\ {\rm (syst)}$, which establishes
\CP violation in the \Bz\ meson system at the 4$\sigma$ level. $B$
lifetime and mixing measurements from a sub-sample of 23 million
$B\Bbar$ pairs are also presented.

\vfill
\begin{center}
Invited talk presented at the \\
International Europhysics Conference on HEP\\
(HEP2001)\\
12--18 July 2001, Budapest, Hungary
\end{center}

\vfill
\begin{center}
{\em Stanford Linear Accelerator Center, Stanford University,
Stanford, CA 94309} \\
\vspace{0.1cm}\hrule\vspace{0.1cm} Work supported by the U.K.
Particle Physics and Astronomy Research Council, Advanced Research
Fellowship GR/L04177 and by Department of Energy contract
DE-AC03-76SF00515.
\end{center}
}

\section{Introduction}
Since its discovery in 1964 in the decays of \KL\
mesons~\cite{KLCP} \CP\ violation has been the subject of many
experiments and the motivation for many theoretical developments
in particle physics. The three-generation Standard Model
accommodates \CP\ violation through the presence of a non-zero
imaginary phase in the Cabibbo-Kobayashi-Maskawa (CKM) quark
mixing matrix~\cite{CKM}. The result presented here establishes
for the first time the existence of \CP\ violation outside the
neutral Kaon system.

The primary goal of the \babar\ experiment at \pep2\ is to perform
stringent tests of the Standard Model by over-constraining the
Unitarity Triangle through \CP\ violation measurements (angles
$\alpha$, $\beta$ and $\gamma$) and the determination of its sides
(\Vub, \Vcb in semileptonic \B\ decays and \Vtd\ in \BzBzb\
mixing).

\section{PEP-II}
The \pep2\ $B$ Factory~\cite{BABARNIM} is an \epem\ colliding beam
storage ring complex at SLAC designed to produce a luminosity of
3x$10^{33} \cm^{-2}s^{-1}$ at a center-of-mass energy of 10.58\gev
(\FourS\ resonance). At the \FourS\ resonance \B\ mesons can only
be produced as {\ensuremath{B^+ {\kern -0.16em B^-}} or coherent
\BzBzb\ pairs. The time evolution of a coherent \BzBzb\ pair is
coupled in such a way that the \CP\ or flavor of one \B\ at decay
time $t_1$ can be described as a function of the other \B\ flavor
at its decay time $t_2$ and the signed time difference $\deltat =
t_1 - t_2$. The machine has asymmetric energy beams (9.0\gev\
electrons on 3.1\gev\ positrons), corresponding to a
center-of-mass boost of $\rm {\beta\gamma}$=0.56. An average
separation of $\rm {\beta\gamma c \tau}\approx$250\mum\ between
the two \B\ meson decay vertices allows the measurement of
time-dependent decay rate asymmetries. \pep2\ has exceeded its
design luminosity by 30\% while \babar, with a logging efficiency
of $>$95\%, has been accumulating data at daily rates up to $260
\invpb$.

\section{\babar}
A detailed description of the detector and its performance can be
found in~\cite{BABARNIM}. The volume within the 1.5T \babar\
superconducting solenoid contains a five layer silicon strip
vertex detector (SVT), a central drift chamber with a helium-based
gas mixture (DCH), a quartz-bar Cherenkov radiation detector
(DIRC) and a CsI(Tl) crystal electromagnetic calorimeter (EMC).
Two layers of cylindrical resistive plate counters (RPCs) are
located between the barrel calorimeter and the magnet cryostat.
The instrumented flux return (IFR) outside the cryostat is
composed of 18 layers of radially increasing thickness steel,
instrumented with 19 layers of planar RPCs in the barrel and 18 in
the endcaps which provide muon and neutral hadron identification.

The SVT has a typical single hit resolution of 15\mum\ in $z$ and
97\% efficiency, while fully reconstructed single \B\ decay vertex
resolution in $z$ is 50\mum. Charged particle tracking using the
SVT and DCH achieves a resolution of $\left( \sigma( p_T / p_T)
\right)^2 =  (0.0015\, p_T)^2 + (0.0045)^2$, where $p_T$ is the
transverse momentum in \gevc. Photons are reconstructed in the
EMC, yielding mass resolutions of 6.9\mevcc\ for \piz\ra\gaga\ and
10\mevcc\ for \KS\ra\piz\piz.

Leptons and hadrons are identified using a combination of
measurements from all the \babar\ components, including energy
loss ${\rm d}E/{\rm d}x$ in the DCH and in the SVT. Electron
identification is mainly based on deposited energy and shower
characteristics in the EMC, while muons are identified in the IFR
and confirmed by their minimum ionizing signal in the EMC.
Excellent kaon identification in the barrel region is provided by
the DIRC, which achieves a $K-\pi$ separation of $>$3.4$\sigma$ in
the range 0.25--3.0\gevc.

\section{Mixing and \CP\ violation with dilepton events}
The mass difference \deltamd\ between the two mass eigenstates of
the neutral $B$ system (sensitive to $|V_{td}|$) is measured by
comparing the rate of neutral $B$ meson pairs decaying with the
same $b$ quark flavor to the rate of decays with the opposite
flavor sign. At the $\FourS$ this time dependent asymmetry gives
direct access to \deltamd:
\begin{equation} A_{\deltamd}(\deltat)= \frac{N(\Bz\Bzb)(\deltat) - (N(\Bz\Bz)(\deltat) +
N(\Bzb\Bzb)(\deltat) )}{N(\Bz\Bzb)(\deltat) + (N(\Bz\Bz)(\deltat)
+ N(\Bzb\Bzb)(\deltat) )} = \cos (\deltamd\cdot\deltat)
\end{equation}
where $\deltat$ is the difference between the two neutral $B$
decay times. The simplest way to determine the $b$ quark flavor of
the decaying neutral $B$  is to use primary leptons as tagging
particles. $A_{\deltamd}(\deltat)$ can be constructed directly
from the rates of ``like-sign'' $(l^+,l^+) + (l^-,l^-)$ and
``unlike-sign'' $(l^+,l^-)$ events. A sample of 100,000 dilepton
events has been selected from 23 million $B\Bbar$ pairs (charged
and neutral). Cascade decays background ($b \rightarrow c
\rightarrow l$) is suppressed by a neural network using kinematic
information of the events. The charged $B$ content of the sample
is extracted from the data together with \deltamd. In the {\em
boost approximation} used in this measurement the decay time
difference is calculated as: $\deltat = \deltaz / c < \beta\gamma
>$, where the small flight path of the \B\ mesons perpendicular to
the z axis is ignored. The \deltaz\ resolution is extracted from
simulation and has been validated by comparisons to real data
control samples (e.g. \jpsi\ decays). The time dependent asymmetry
is shown in Fig.~\ref{fig:dileptons}. The preliminary result is:
\begin{equation}
\deltamd = 0.499 \pm 0.010 ({\rm stat.}) \pm 0.012 ({\rm syst.})
\hbar \ps^{-1}
\end{equation}

\CP\ and $T$ violation in \BzBzb\ mixing can be probed in the same
way as in \KzKzb\ mixing~\cite{cplearT}. Its magnitude is given by
the parameter $\varepsilon$:
\begin{equation} A_T(|\deltat|)=\frac{Nb(l^+,l^+) - Nb(l^-,l^-)}
{ Nb(l^+,l^+) + Nb(l^-,l^-)} \approx
\frac{4Re(\varepsilon)}{1+|\varepsilon|^2}
\end{equation}
Using our dilepton sample we obtain the preliminary result:
\begin{equation}
\frac{4Re(\varepsilon)}{1+|\varepsilon|^2} = (1.2 \pm 2.9 ({\rm
stat.}) \pm 3.6 ({\rm syst.}) ) \times 10^{-3}
\end{equation}
which is the most precise measurement of this quantity to date.
The $A_T$ distribution is also shown in Fig.~\ref{fig:dileptons}.

\begin{figure}[htbp]
\begin{center}
 \mbox{\epsfig{file=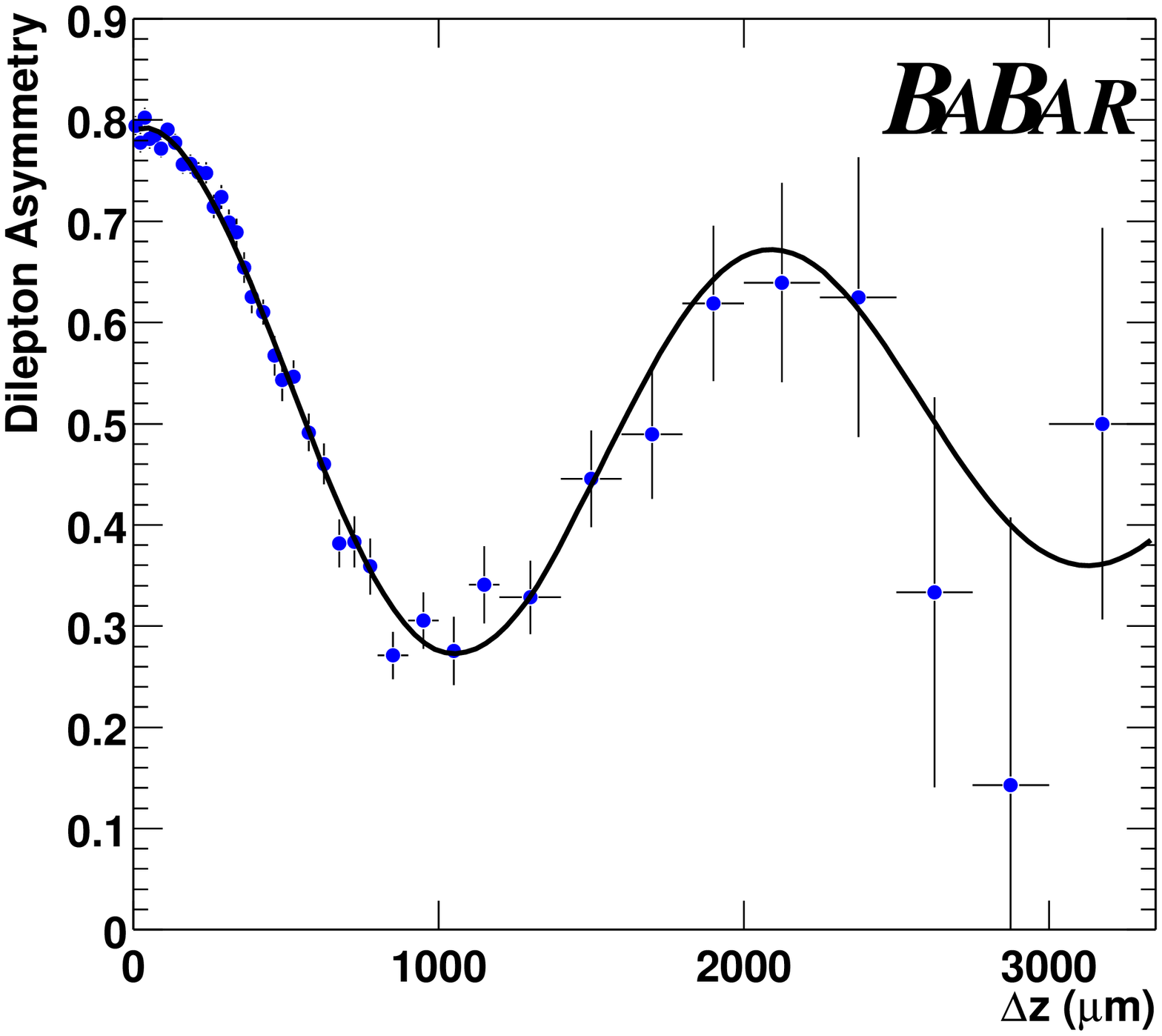,height=150pt}}
 \mbox{\epsfig{file=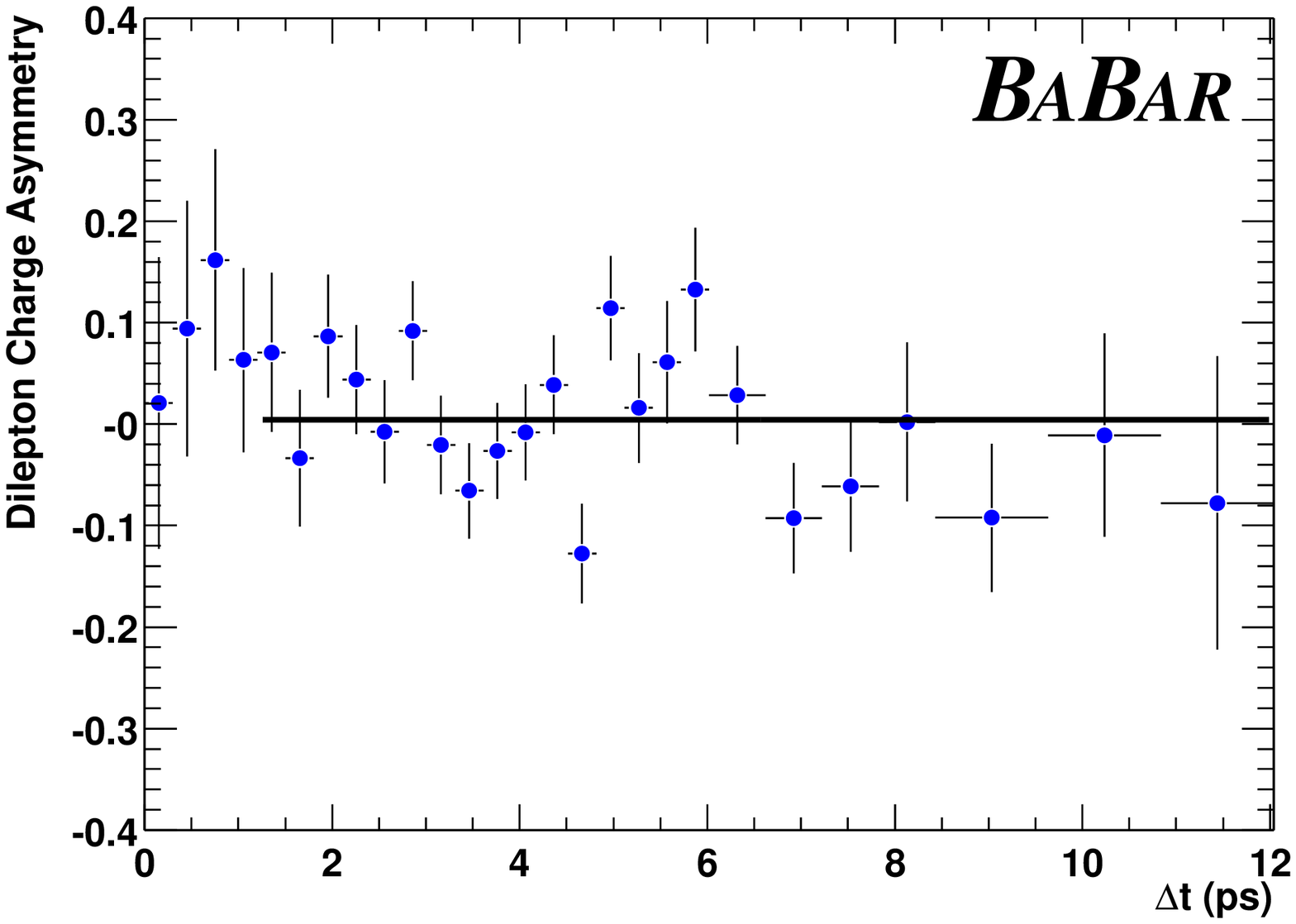,height=150pt}}
\end{center}
\caption{Left: The time dependent asymmetry between unmixed and
mixed dilepton events. Right: The asymmetry $A_T$. The fit results
are superimposed on both asymmetry plots.} \label{fig:dileptons}
\end{figure}

\section{Time-dependent measurements with exclusively reconstructed \B\ events}
High purity \B\ event samples are obtained when the hadronic decay
of one \B\ meson is fully reconstructed. In such events the
kinematics and decay vertex, as well as the flavor or \CP\ content
of the exclusively reconstructed \B\ are fully determined. The
events where in addition the decay vertex of the other \B\ meson
can be determined are used for lifetime measurements. The \BzBzb\
events where both the decay vertex and flavor (at decay time) of
the other \B\ can be determined are used for mixing or \CP\
violation measurements.

\subsection{Exclusive \B\ reconstruction}
\Bz mesons are fully reconstructed in hadronic modes of (a)
definite flavor: \Bz $\to D^{(*)-} \pi^+$, $D^{(*)-} \rho^+$,
$D^{(*)-} a_1^+$, $\jpsi \Kstarz(\Kstarz\to K^+ \pi^-)$ and (b)
known \CP\ content: \Bz $\to \jpsi\KS$, $\psitwos\KS$, $\jpsi\KL$,
$\chicone \KS$, $\jpsi\Kstarz(\Kstarz\to\KS\piz)$. In the
following the two samples are referred to as $B_{{\rm flav}}$ and
$B_{\CP}$ respectively. \Bpm mesons are reconstructed in the
hadronic modes \Bub $\to \Dz \pim$, $\Dstarz \pim$, $\jpsi K^-$,
$\psitwos K^-$ (throughout this paper conjugate modes are
implied).

The selections have been optimized for signal significance, using
on-peak, off-peak and simulated data. Charged particle
identification, mass (or mass difference) and vertex constraints
are used wherever applicable. The signal for each decay mode is
identified in the two-dimensional distribution of the kinematical
variables $\Delta E$ and \mes: $\Delta E=E^*_{\rm rec}-E^*_b$ is
the difference between the \B\ candidate energy and the beam
energy and $\mes=\sqrt{E^{*2}_b - \mbox{\boldmath $p$}^{*2}_{\rm
rec}}$ is the mass of a particle with a reconstructed momentum
$\mbox{\boldmath $p$}^*_{\rm rec} = \sum_i \mbox{\boldmath
$p$}^*_i$ assumed to have the beam energy, as is the case for a
true \B\ meson. In events with several \B\ candidates only the one
with the smallest $\Delta E$ is considered. The $\Delta E$ and
\mes variables have minimal correlation. The resolution in \mes is
$\approx$3\mevcc, dominated by the beam energy spread. The
resolution in $\Delta E$ is mode dependent and varies in the range
of 12--40\mev. For each mode a rectangular signal region is
defined by the three standard deviation bands in \mes ($5.27 <
\mes < 5.29$\gevcc) and $\Delta E$ (mode dependent interval). The
composition of each sample is determined by fitting the \mes
distribution for candidates within the signal region in $\Delta E$
to the sum of a single Gaussian representing the signal and a
background function introduced by the ARGUS
collaboration~\cite{bib:argusfunction}.

\subsection{\deltat\ calculation and resolution}
Since no stable charged particle emerges from the \FourS\ decay
point, the production point of the \B\ mesons and thus their
individual decay times cannot be determined. However the decay
time difference \deltat\ between the two is sufficient for the
description of a coherent \B\ meson pair (decay length difference
technique).

The difference of the proper decay times of the \B\ mesons
\deltat=$t_1-t_2$ is determined from the separation along the
boost direction \deltaz=$z_1-z_2$, including an event-by-event
correction for the direction of the $B$ with respect to the $z$
direction in the $\FourS$ frame. $z_1$ is determined from the
charged tracks that constitute the exclusively reconstructed $B_1$
candidate. The other $B$ vertex is determined by fitting the
tracks not belonging to the $B_1$ candidate to a common vertex.
Tracks from photon conversions are removed. Pairs of tracks
compatible with the decay of a long lived \Kz\ or $\Lambda$ are
replaced by the parent neutral pseudotrack. To reduce the bias in
the forward $z$ direction from charm decay products, the track
with the largest contribution to the vertex $\chi^2$, if above 6,
is removed and the fit is iterated until no track fulfills this
condition. Knowledge of the beam spot location and beam direction
is incorporated in the $B_2$ vertex determination through the
addition of a pseudotrack to its vertex, computed from the $B_1$
vertex and three-momentum, the beam spot (with a vertical size of
10\mum ) and the \FourS momentum. The $\deltaz$ reconstruction
efficiency is 97\%. For 99\% of the reconstructed vertices the
r.m.s. \deltaz resolution measured in data is 180\mum, dominated
by the $z_2$ vertex. An event is accepted if it has converged fits
for the two \B\ vertices, an error of less than 400\mum on
\deltaz, and a measured $\vert \deltat \vert < 20 \ps$.

The modelling of the resolution function $\cal{R}$ is a crucial
element of all time-dependent measurements. In the case of
\B~lifetime measurements, studies both on simulated and real data
have shown that adding a zero-mean Gaussian distribution and its
convolution with a decay exponential provides the best compromise
between different sources of uncertainties. The width of the
Gaussian is scaled by the per-event \deltat\ error, derived from
the vertex fits. A zero-mean Gaussian distribution with width
fixed at 10\ps is used to describe outliers (less than 1\% of
events with incorrectly reconstructed vertices).

In the mixing and \CP\ violation measurements the time resolution
is described by the sum of three Gaussian distributions (core,
tail and outliers) with different means. In the \CP\ fit the means
are modeled to be proportional to the per-event \deltat\ error,
which is correlated with the weight that the daughters of
long-lived charm particles have in the tag vertex reconstruction.
The core and tail widths are scaled by the per-event \deltat\
error. The outlier width is fixed to 8\ps.

\subsection{Flavor tagging}
For flavor tagging we exploit information from the incompletely
reconstructed other $B$ decay in the event. Each event is assigned
to one of four hierarchical, mutually exclusive tagging categories
or excluded from further analysis. The {\tt Lepton} and {\tt Kaon}
categories contain events with high momentum leptons from
semileptonic \B\ decays or with kaons whose charge is correlated
with the flavor of the decaying $b$ quark ({\it e.g.} a positive
lepton or kaon yields a \Bz tag). The {\tt NT1} and {\tt NT2}
categories are based on a neural network algorithm whose tagging
power arises primarily from soft pions from $D^{*+}$ decays and
from recovering unidentified isolated primary leptons.

The figure of merit for tagging is the effective tagging
efficiency $Q_i = \eps_i (1-2\mistag_i)^2$, where $\eps_i$ is the
fraction of events with a reconstructed tag vertex that are
assigned to the $i^{th}$ category and $\mistag_i$ is the mistag
fraction for the same category. The statistical error on \stwob is
proportional to $1/\sqrt{Q}$, where $Q=\sum Q_i$. The efficiencies
and mistag fractions for the four tagging categories are measured
from data and summarized in Table~\ref{tab:mistag}.

\begin{table}[hb]
\begin{center}
\begin{tabular}{|l|c|c|c|c|} \hline
Tagging Category & $\varepsilon$ (\%) & $\mistag$ (\%) &
$\Delta\mistag$ (\%)& $Q$ (\%)
\\ \hline \hline
{\tt Lepton} & $10.9\pm0.3$ & $8.9\pm1.3$  & $0.9\pm2.2$  & $7.4\pm0.5$  \\
{\tt Kaon}   & $35.8\pm0.5$ & $17.6\pm1.0$ & $-1.9\pm1.5$ & $15.0\pm0.9$  \\
{\tt NT1}    & $7.8\pm0.3$  & $22.0\pm2.1$ & $5.6\pm3.2$  & $2.5\pm0.4$  \\
{\tt NT2}    & $13.8\pm0.3$ & $35.1\pm1.9$ & $-5.9\pm2.7$ &
$1.2\pm0.3$
\\ \hline \hline
all          & $68.4\pm0.7$ &       &           &  $26.1\pm1.2$ \\
\hline
\end{tabular}
\end{center}
\caption { Average mistag fractions $\mistag_i$ and mistag
differences $\Delta\mistag_i=\mistag_i(\Bz)-\mistag_i(\Bzb)$
extracted for each tagging category $i$ from the
maximum-likelihood fit to the time distribution for the
fully-reconstructed \Bz\ sample ($B_{\rm flav}$+$B_{\CP}$).
Uncertainties are statistical only.} \label{tab:mistag}
\end{table}

\subsection{\B\ lifetime and mixing measurements}
The results presented here have been obtained from a sample of
approximately 23 million $B\overline{B}$ pairs collected by
\babar\ between October 1999 and October 2000. Samples of
$\approx$6000 \Bz\ and $\approx$6300 \Bu\ signal events have been
selected with background contamination of less than 10\%. The
results of a fit with a Gaussian signal distribution and an ARGUS
background function~\cite{bib:argusfunction} are superimposed on
the \mes\ distribution of the final sample in
Fig.~\ref{fig:breco}. The $B$ meson lifetimes are extracted from
unbinned maximum likelihood fits to the \deltat\ distributions,
also shown in Fig.~\ref{fig:breco}. We obtain:
\begin{eqnarray}
\small
 \tau_{\Bz} &=& 1.546 \pm 0.032\mbox{ (stat)} \pm 0.022\mbox{ (syst)} \mbox{ ps}        \nonumber \\
 \tau_{\Bu} &=& 1.673 \pm 0.032\mbox{ (stat)} \pm 0.023\mbox{ (syst)} \mbox{ ps}    \nonumber \\
 \tau_{\Bu}/\tau_{\Bz} &=& 1.082 \pm 0.026\mbox{ (stat)} \pm 0.012\mbox{ (syst)} \nonumber
\normalsize
\end{eqnarray}

\begin{figure}[htbp]
\begin{center}
 \mbox{\epsfig{file=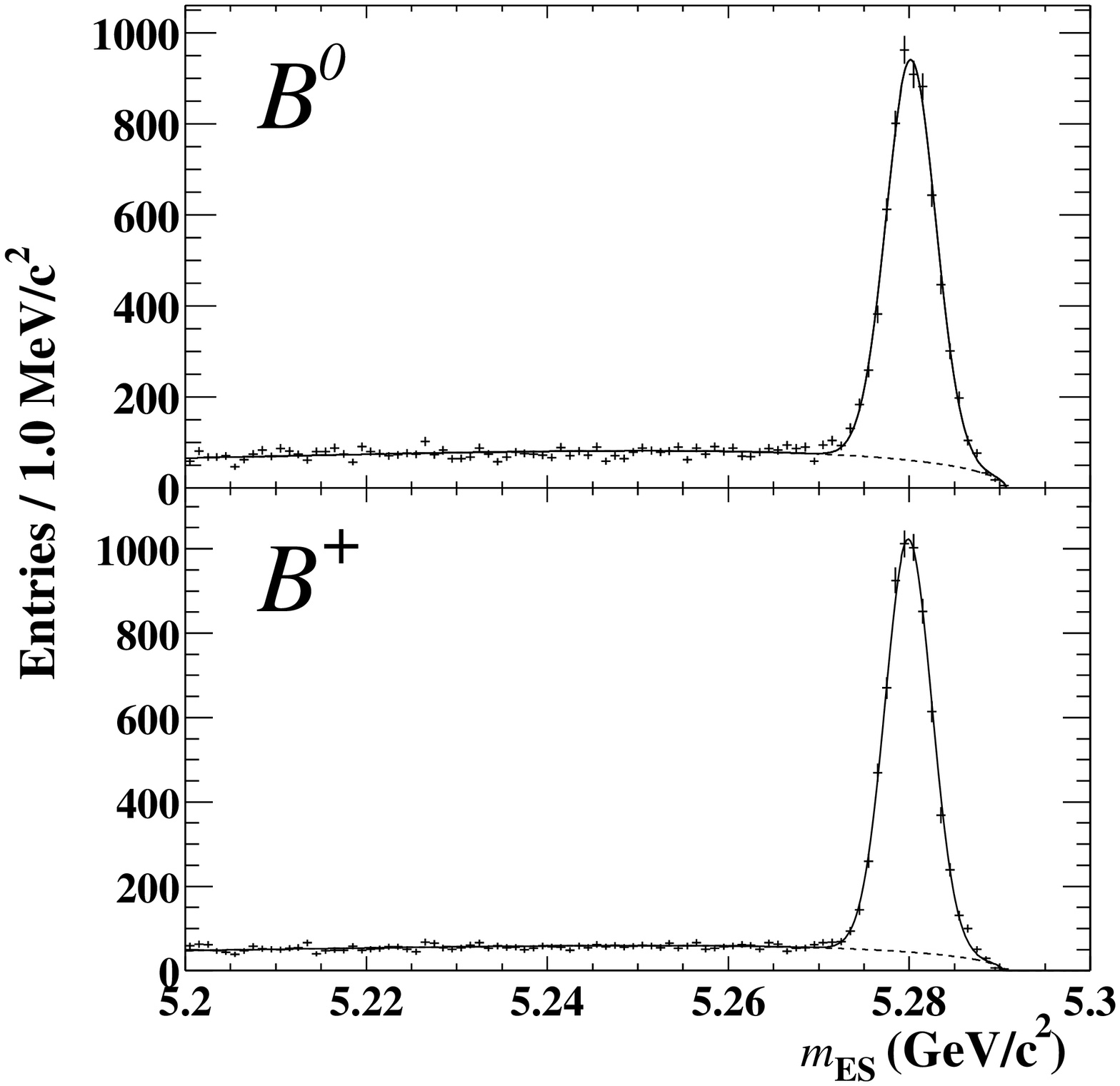,height=200pt}}
 \mbox{\epsfig{file=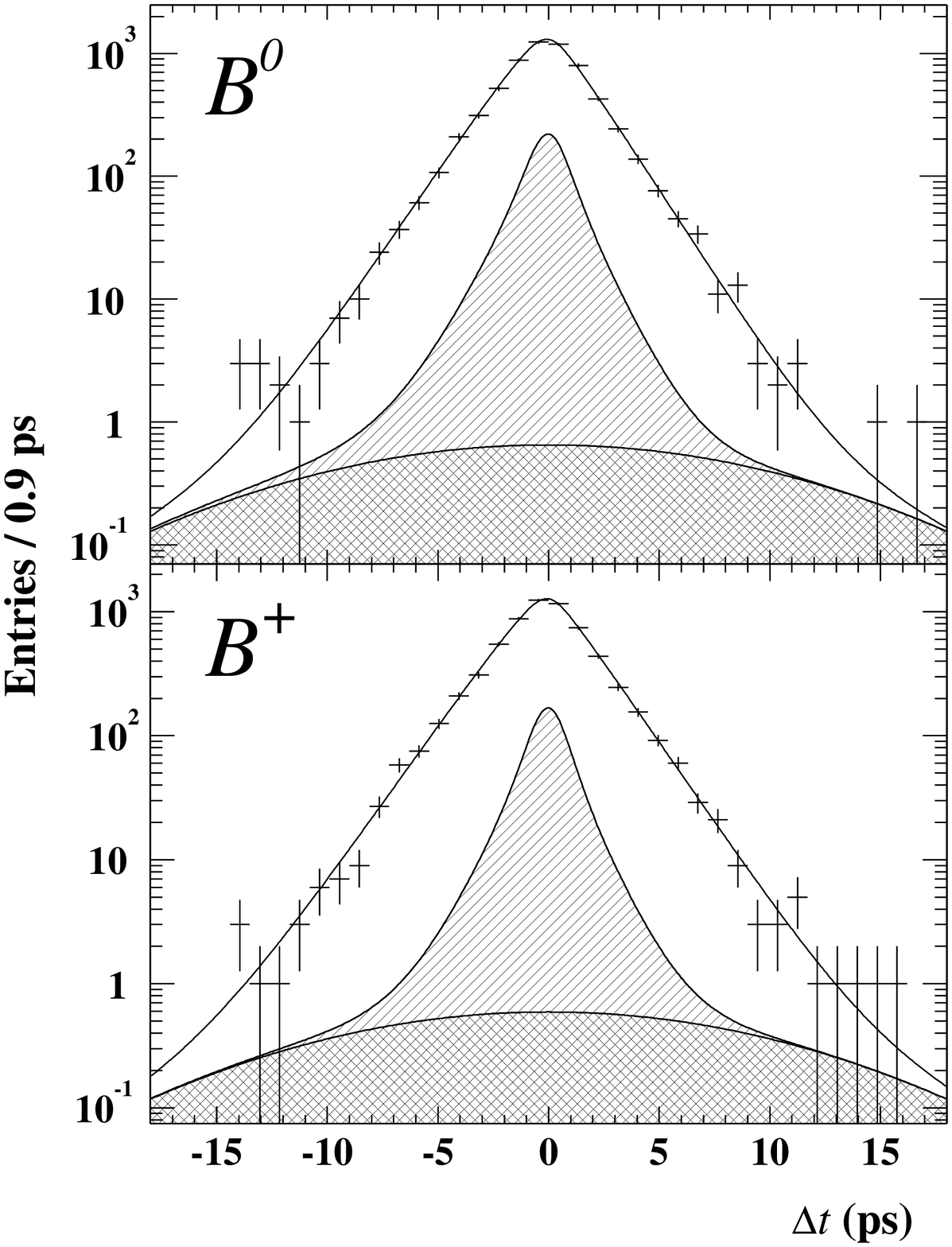,height=200pt}}
\end{center}
\caption{Left: \mes\ distributions of the selected neutral (top)
and charged (bottom) $B$ candidates. Right: \deltat\ distribution
for the \Bz\ (top) and \Bu\ (bottom) events within $2 \sigma$ of
the \B mass in \mes\ with superimposed fit results. The
single-hatched areas are the background components and the
cross-hatched areas represent the outlier contributions.}
\label{fig:breco}
\end{figure}

Neutral $B$ meson pairs, produced as \BzBzb\ decay either as
\BzBzb\ (unmixed) or \Bz\Bz(\Bzb\Bzb) (mixed), allowing to observe
\BzBzb\ mixing and measure \deltamd. A sample of $\approx$4500
such events where the flavor of the second $B$ at the time of its
decay has been tagged has been obtained. The \deltat\
distributions and the mixing oscillation are shown in
Fig.~\ref{fig:exclmix}. From an unbinned maximum likelihood fit we
obtain the preliminary result:
\begin{equation}
\deltamd = 0.519 \pm 0.020 ({\rm stat.}) \pm 0.016 ({\rm syst.})
\hbar \ps^{-1}
\end{equation}

\begin{figure}[htbp]
\begin{center}
 \mbox{\epsfig{file=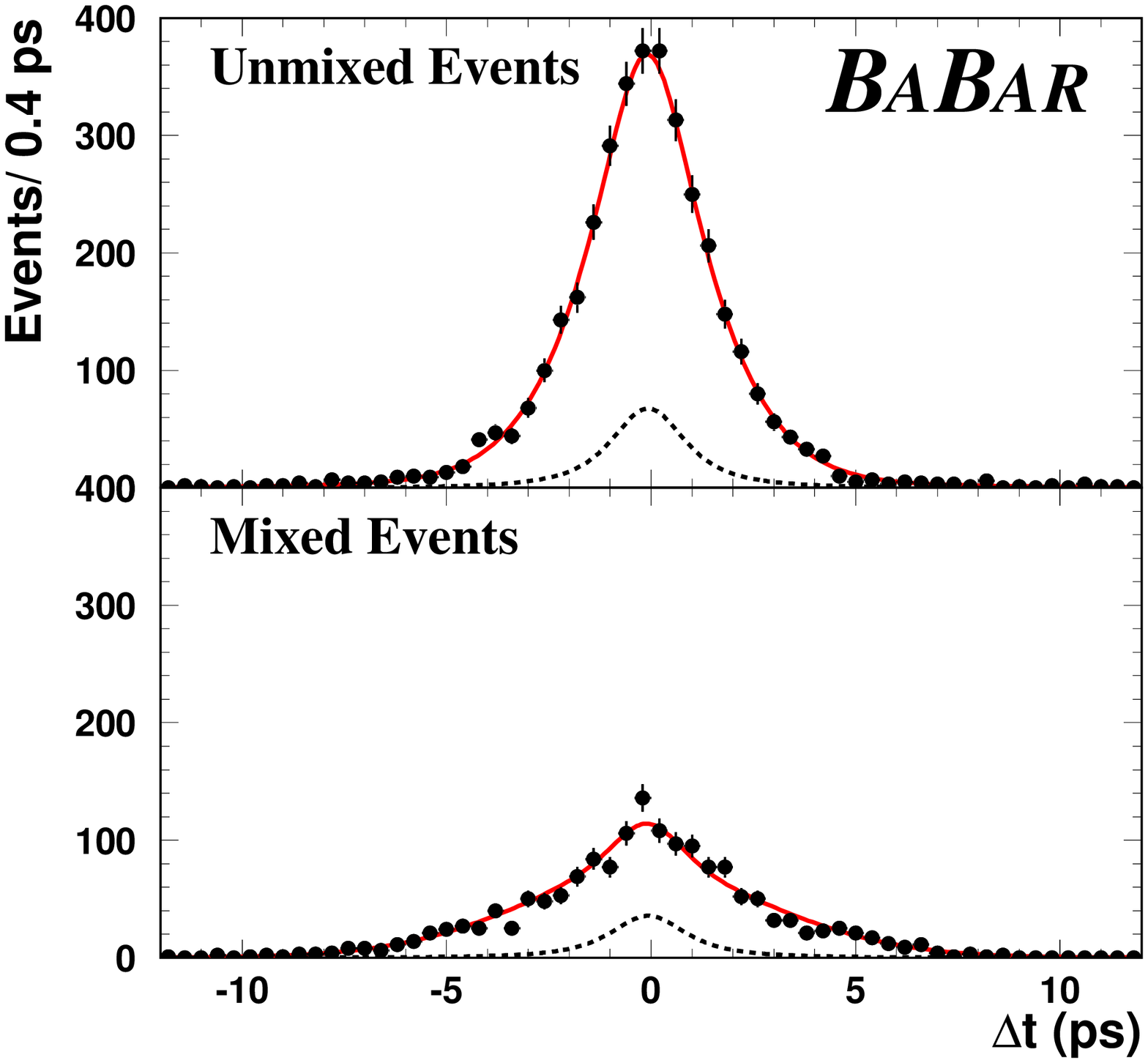,height=150pt}}
 \mbox{\epsfig{file=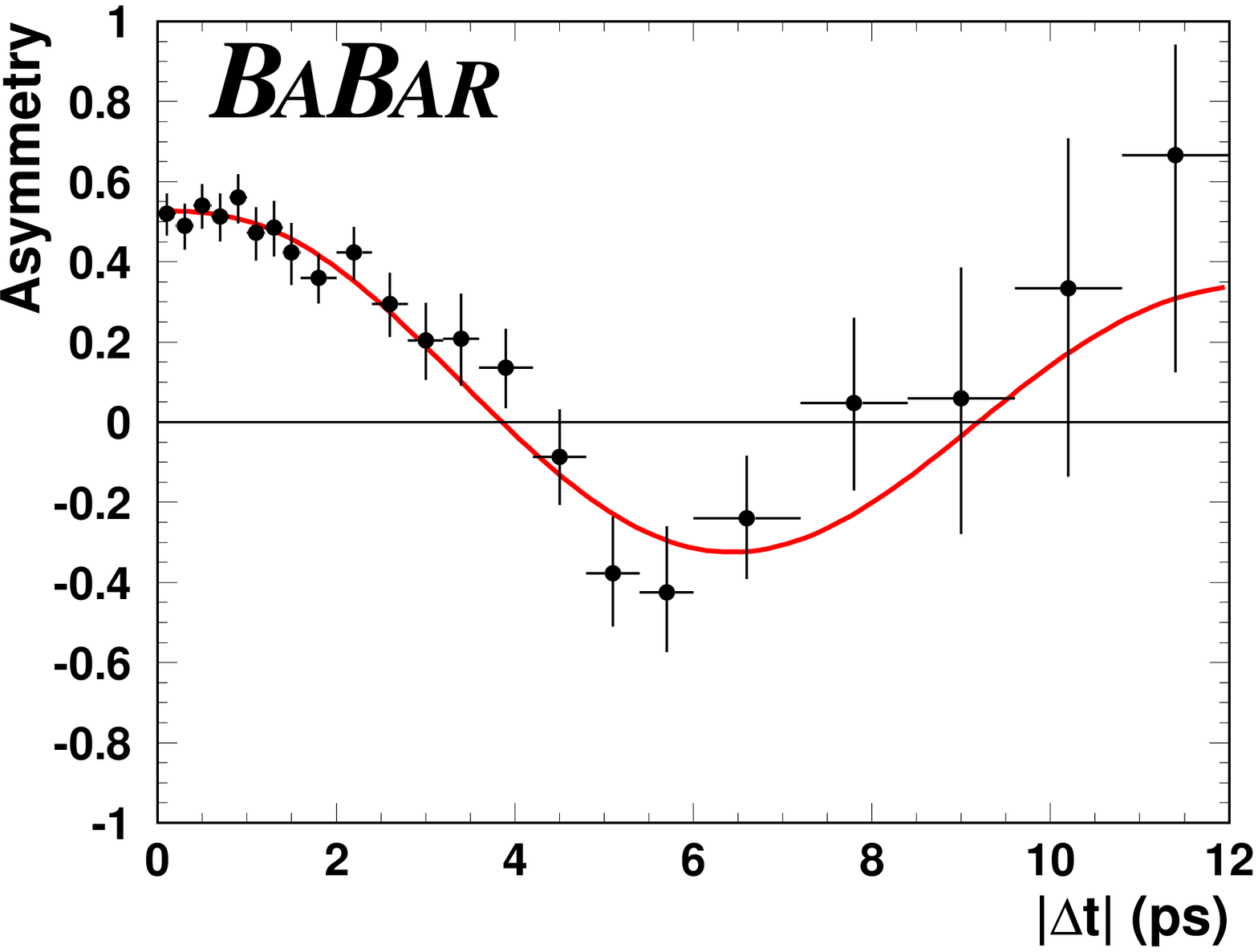,height=150pt}}
\end{center}
\caption{Left: \deltat\ distributions of the selected neutral
unmixed (top) and mixed (bottom) events. Right: The time dependent
asymmetry between unmixed and mixed events. The fit results are
shown superimposed in all distributions. The dashed curves on the
\deltat\ plots indicate the background contributions.}
\label{fig:exclmix}
\end{figure}

\section{\stwob\ and the observation of \CP\ violation}
The data set of 32 million $B\Bbar$ pairs collected between
October 1999 and May 2001 has been used to fully reconstruct a
sample $B_{\CP}$ of neutral $B$ mesons decaying to the $\jpsi\KS$,
$\psitwos\KS$, $\jpsi\KL$, $\chicone\KS$, and $\jpsi\Kstarz
(\Kstarz\to\KS\piz ) $ final states. The last two modes have been
added since our first \stwob\ publication~\cite{BABARPRL}. There
are several other significant changes in the analysis.
Improvements in track and \KS\ reconstruction efficiency in 2001
data produce a $\approx$30\% increase in the yields for a given
luminosity. Better alignment of the tracking systems in 2001 data
and improvements in the tagging vertex reconstruction algorithm
increase the sensitivity of the measurement by an additional 10\%.
Optimization of the  $\jpsi\KL$ selection increases the purity of
this sample. The final $B_{\CP}$ sample contains about 640 signal
events and, with all the improvements, the statistical power of
the analysis is almost doubled with respect to  that of
Ref.~\cite{BABARPRL}. A sample $B_{flav}$ of 7591 fully
reconstructed \Bz\ events in definite flavor eigenstates has been
used to determine the tagging performance and a sample of 6814
fully reconstructed charged $B$ has been used for validations.
Details of our samples are shown in Table~\ref{tab:cpsamples}. The
samples are shown in Fig.~\ref{fig:cpsamples}.

\begin{figure}[!]
\begin{center}
\epsfig{file=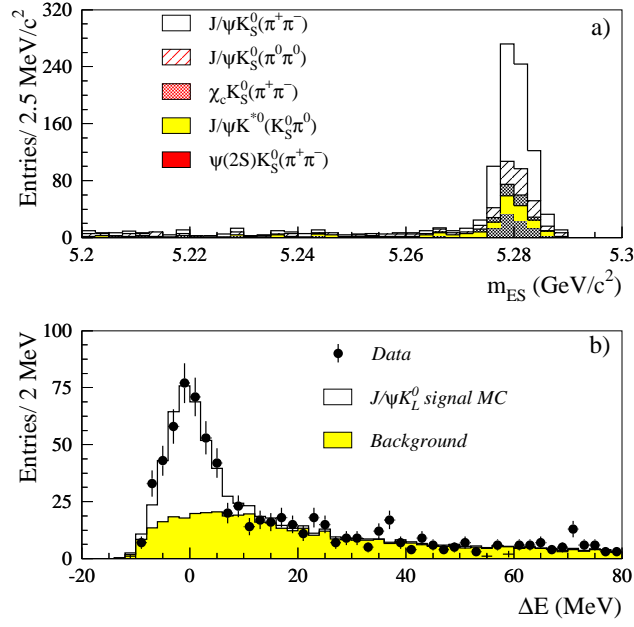,height=230pt} \caption{ a)
Distribution of \mes\ for $B_{\CP}$ candidates having a \KS in the
final state; b) distribution of $\Delta E$ for $\jpsi\KL$
candidates.} \label{fig:cpsamples}
\end{center}
\end{figure}

\begin{table}[!htb]
\caption{ Number of tagged events, signal purity and result of
fitting for \CP\ asymmetries in the full \CP sample and in various
subsamples, as well as in the $B_{\rm flav}$ and charged $B$
control samples. Errors are statistical only.}
\label{tab:cpsamples}
\begin{center}
\begin{tabular}{c c c c } \hline \hline
Sample                    & $N_{\rm tag}$   & Purity (\%) & \stwob
\\ \hline
$\jpsi\KS$,$\psitwos\KS$,$\chicone\KS$   & $480$        & $96$       &  0.56 $\pm$ 0.15   \\
$\jpsi \KL$ $(\eta_f=+1)$                & $273$        & $51$       &  0.70 $\pm$ 0.34   \\
$\jpsi\Kstarz ,\Kstarz \to \KS\piz       $& $50$         & $74$       &  0.82 $\pm$ 1.00  \\
\hline
 Full \CP\ sample                        & $803$        & $80$       &  0.59 $\pm$ 0.14   \\
\hline \hline
\multicolumn{4}{l}{$\jpsi\KS$, $\psitwos\KS$, $\chicone\KS$  only  $(\eta_f=-1)$ }  \\
\hline
$\ \jpsi \KS$ ($\KS \to \pi^+ \pi^-$)    & $316$        & $98$       &  0.45 $\pm$ 0.18  \\
$\ \jpsi \KS$ ($\KS \to \pi^0 \pi^0$)    & $64$         & $94$       &  0.70 $\pm$ 0.50  \\
$\ \psi(2S) \KS$ ($\KS \to \pi^+ \pi^-$) & $67$         & $98$       &  0.47 $\pm$ 0.42   \\
$\ \chicone \KS $                        & $33$         & $97$       &  2.59 $\pm$ $^{0.55}_{0.67}$ \\
\hline
$\ $ {\tt Lepton} tags                   & $74$         &  $100$     &  0.54 $\pm$ 0.29   \\
$\ $ {\tt Kaon} tags                     & $271$        &  $98$      &  0.59 $\pm$ 0.20    \\
$\ $ {\tt NT1} tags                      & $46$         &  $97$      &  0.67 $\pm$ 0.45    \\
$\ $ {\tt NT2} tags                      & $89$         &  $95$      &  0.10 $\pm$ 0.74   \\
\hline
$\ $ \Bz\ tags                           & $234$        &  $98$      &  0.50 $\pm$ 0.22     \\
$\ $ \Bzb\ tags                          & $246$        &  $97$      &  0.61 $\pm$ 0.22     \\
\hline\hline
$B_{\rm flav}$ non-\CP sample            & $7591$       & $86$       &  0.02 $\pm$ 0.04     \\
\hline
Charged $B$ non-\CP sample        & $6814$       & $86$       &  0.03 $\pm$ 0.04     \\
\hline \hline
\end{tabular}
\end{center}
\end{table}

We examine each of the events in the $B_{\CP}$ sample for evidence
that the other neutral $B$ meson decayed as a \Bz or a \Bzb
(flavor tag). The decay-time distributions for events with a \Bz
or a \Bzb tag can be expressed in terms of a complex parameter
$\lambda$ that depends on both \BzBzb mixing and on the amplitudes
describing \Bzb and \Bz decay to a common final state
$f$~\cite{lambda}. The distribution ${\rm f}_+({\rm f}_-)$ of the
decay rate when the tagging meson is a $\Bz (\Bzb)$ is given by
{\small{
\begin{eqnarray}
{\rm f}_\pm(\, \deltat) = {\frac{{\rm e}^{{- \left| \deltat
\right|}/\tau_{\Bz} }}{2\tau_{\Bz} (1+|\lambda|^2) }}  \times
\left[ \ {\frac{1 + |\lambda|^2}{2}} \pm {\ \mathop{\cal I\mkern
-2.0mu\mit m}} \lambda  \sin{( \Delta m_{B^0}  \deltat )} \mp {
\frac{1  - |\lambda|^2 } {2} }
  \cos{( \Delta m_{B^0}  \deltat) }
  \right] \label{eq:timedist}
\end{eqnarray}
}} where $\deltat=t_{CP}-t_{\rm tag}$ is the time between the two
\B decays, $\tau_{\Bz}$ is the \Bz lifetime and $\Delta m_{B^0}$
is the mass difference determined from \BzBzb
mixing~\cite{PDG2000}. The first oscillatory term in
Eq.~\ref{eq:timedist} is due to interference between direct decay
and decay after mixing, and the second term is due to direct \CP\
violation. A difference between the \Bz and \Bzb \deltat
distributions or a \deltat asymmetry for either flavor tag is
evidence for \CP violation.

In the Standard Model $\lambda=\eta_f e^{-2i\beta}$ for
charmonium-containing $b\to\ccbar s$ decays, $\eta_f$ is the \CP
eigenvalue of the state $f$ and $\beta = \arg \left[\, -V_{\rm
cd}^{}V_{\rm cb}^* / V_{\rm td}^{}V_{\rm tb}^*\, \right]$ is an
angle of the Unitarity Triangle. Thus, the time-dependent
\CP-violating asymmetry is
\begin{eqnarray}
A_{\CP}(\deltat) &\equiv&  \frac{ {\rm f}_+(\deltat)  -  {\rm
f}_-(\deltat) } { {\rm f}_+(\deltat) + {\rm f}_-(\deltat) } =
-\eta_f \stwob \sin{ (\Delta m_{B^0} \, \deltat )} ,
\label{eq:asymmetry}
\end{eqnarray}
where $\eta_f=-1$ for $\jpsi\KS$, $\psitwos\KS$ and $\chicone \KS$
and $+1$ for $\jpsi\KL$. Due to the presence of even (L=0, 2) and
odd (L=1) orbital angular momenta in the $\jpsi\Kstarz
(\Kstarz\to\KS\piz) $ system, there are \CP-even and \CP-odd
contributions to the decay rate. When the angular information in
the decay is ignored, the measured \CP\ asymmetry in
$\jpsi\Kstarz$ is reduced by a dilution factor $D_{\perp} =
1-2R_{\perp}$, where $R_{\perp}$ is the fraction of the L=1
component. We have measured $R_{\perp} = (16 \pm 3.5)\% $
~\cite{BABARTRANS} which, after acceptance corrections, leads to
an effective $\eta_f = 0.65 \pm 0.07$ for the $\jpsi\Kstarz$ mode.

The $B_{\CP}$ and $B_{flav}$ samples are used together in the
unbinned maximum likelihood fit for the extraction of \stwob. A
total of 45 parameters are varied in the fit, including \stwob
(1), the average mistag fraction $\mistag$ and the difference
$\Delta\mistag$ between \Bz\ and \Bzb\ mistags for each tagging
category (8), parameters for the signal \deltat resolution (16),
and parameters for background time dependence (9), \deltat
resolution (3) and mistag fractions (8). The determination of the
mistag fractions and signal \deltat resolution function is
dominated by the large $B_{\rm flav}$ sample. Background
parameters are governed by events with $\mes < 5.27\gevcc$. As a
result, the largest correlation between \stwob\ and any linear
combination of the other free parameters is only 0.13. We fix
$\tau_{\Bz}=1.548\ps$ and $\Delta
m_{B^0}=0.472\,\hbar\ps^{-1}$~\cite{PDG2000}. The value of \stwob\
and the \CP\ asymmetry in the \deltat\ distribution were hidden
(blind analysis) following our previous
publication~\cite{BABARPRL}, until the event selection was
optimized and all other aspects of the present analysis were
complete.

Figure~\ref{fig:asymlike} shows the $\deltat$ distributions and
${A}_{\CP}$ as a function of \deltat overlaid with the likelihood
fit result for the $\eta_f = -1$ and $\eta_f = +1$ samples. The
probability of obtaining a lower likelihood, evaluated with a
parameterized simulation of a large number of data-sized
experiments, is 27\%. Our result is:
\begin{eqnarray}
\stwob=0.59 \pm 0.14\ \stat \pm 0.05\ \syst. \nonumber
\end{eqnarray}
Repeating the fit with all parameters except \stwob\ fixed to
their values at the global maximum likelihood, we attribute a
total contribution in quadrature of 0.02 to the error on \stwob\
due to the combined statistical uncertainties in mistag fractions,
\deltat\ resolution and background parameters. The dominant
sources of systematic error are the parameterization of the
\deltat\ resolution function (0.03), due in part to residual
uncertainties in SVT alignment, possible differences in the mistag
fractions between the $B_{\CP}$ and $B_{\rm flav}$ samples (0.03),
and uncertainties in the level, composition, and \CP\ asymmetry of
the background in the selected \CP events (0.02). The systematic
errors from uncertainties in $\Delta m_{\Bz}$ and $\tau_{\Bz}$ and
from the parameterization of the background in the $B_{\rm flav}$
sample are small; an increase of $0.02\,\hbar\ps^{-1}$ in the
value for $\Delta m_{\Bz}$ decreases \stwob\ by 0.015.

\begin{figure}[t!]
\begin{center}
\epsfig{file=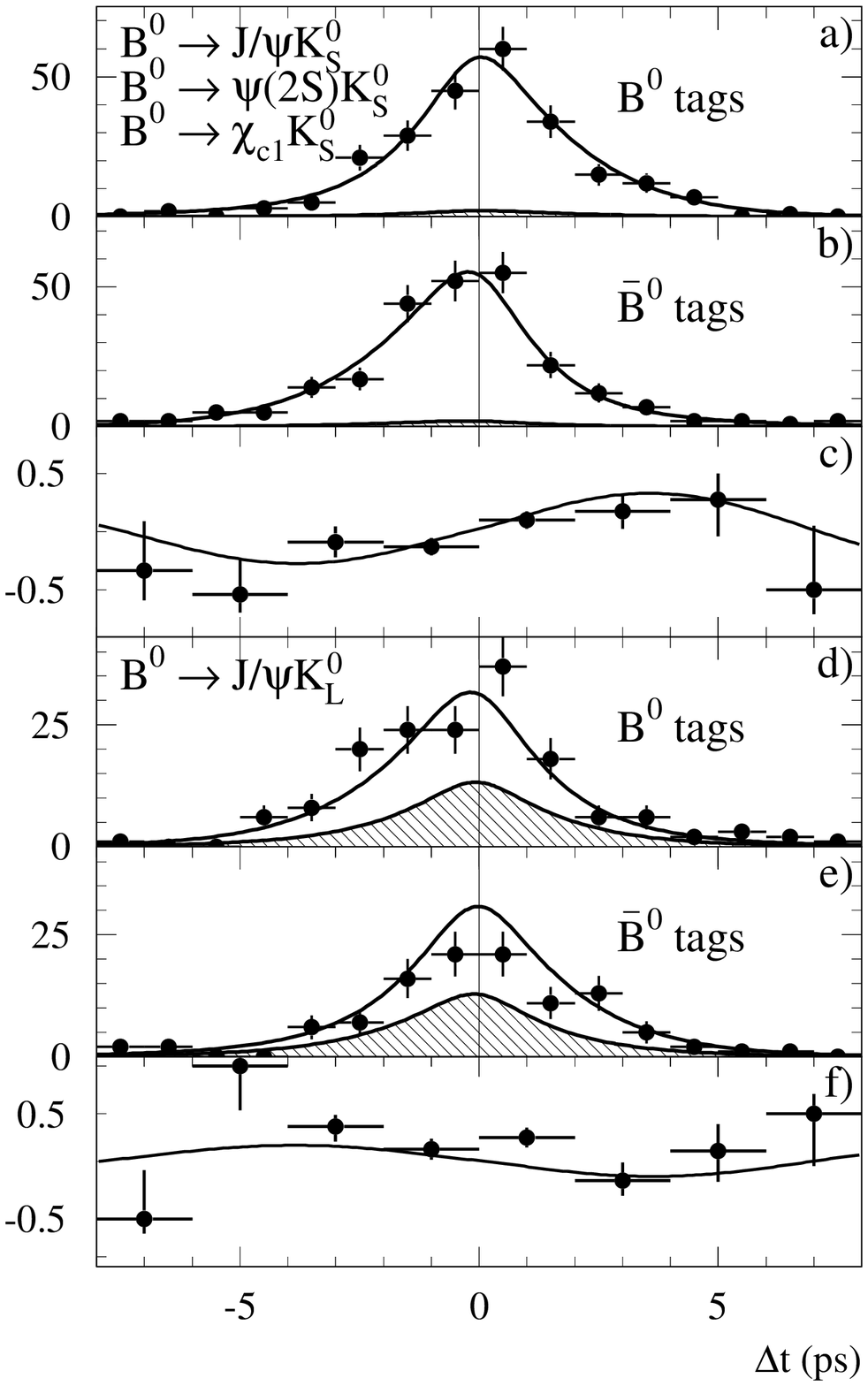, height=10cm} \caption{Number of
$\eta_f=-1$ candidates ($\jpsi\KS$, $\psitwos\KS$, and
$\chicone\KS $) in the signal region  a) with a \Bz tag $N_{\Bz}$
and b) with a \Bzb tag $N_{\Bzb}$, and c) the asymmetry
$(N_{\Bz}-N_{\Bzb})/(N_{\Bz}+N_{\Bzb})$, as functions of \deltat .
The solid curves represent the result of the combined fit to all
selected \CP events; the shaded regions represent the background
contributions. Figures d)--f) contain the corresponding
information for the $\eta_f=+1$ mode $(\jpsi\KL)$. The likelihood
is normalized to the total number of \Bz and \Bzb tags. The value
of \stwob is independent of the individual normalizations and
therefore of the difference between the number of \Bz and \Bzb
tags. } \label{fig:asymlike}
\end{center}
\end{figure}

The large sample of reconstructed events allows a number of
consistency checks, including separation of the data by decay
mode, tagging category and $B_{\rm tag}$ flavor. The results of
fits to these subsamples are shown in Table~\ref{tab:cpsamples}.
Table~\ref{tab:cpsamples} also shows results of fits to the
samples of non-\CP decay modes, where no statistically significant
asymmetry is found. Performing the current analysis on the
previously published data sample and decay modes yields a value of
\stwob=0.32$\pm$0.18,  consistent with the published
value~\cite{BABARPRL}. For only these decay modes, the year 2001
data yield \stwob=0.83$\pm$0.23, consistent with the 1999-2000
results at the 1.8$\sigma$ level; for the $\jpsi\KS
(\KS\to\pip\pim)$ channel the consistency is at the 1.4$\sigma$
level.

If $\vert\lambda\vert$ is allowed to float in the fit to the
$\eta_f=-1$ sample, which has high purity and requires minimal
assumptions on the effect of backgrounds, the value obtained is
$\vert\lambda\vert = 0.93 \pm 0.09\ \rm{\stat} \pm 0.03\
\rm{\syst}$. The sources of the systematic error in this
measurement are the same as in the \stwob analysis. In this fit,
the coefficient of the $\sin{( \Delta m_{B^0}  \deltat )}$ term in
Eq.~\ref{eq:timedist} is measured to be $0.56\pm 0.15$ (\rm{stat})
in agreement with Table~\ref{tab:cpsamples}.

The measurement of $\stwob=0.59 \pm 0.14\ \stat \pm 0.05\ \syst $
establishes \CP violation in the \Bz meson system at the
$4.1\sigma$ level. This significance is computed from the sum in
quadrature of the statistical and additive systematic errors. The
probability of obtaining this value or higher in the absence of
\CP violation is less than $3 \times 10^{-5}$. This direct
measurement is consistent with the range implied by measurements
and theoretical estimates of the magnitudes of CKM matrix
elements~\cite{CKMconstraints}.

\section{Conclusions and prospects}
37 years after the discovery of \CP violation in the Kaon system,
\babar\ has established \CP violation in the $B$ system with the
measurement:
\begin{eqnarray}
\stwob=0.59 \pm 0.14\ \stat \pm 0.05\ \syst. \nonumber
\end{eqnarray}
By the summer of 2002 \babar\ will have a data sample of more than
100 million $B\Bbar$ pairs, bringing the precision on \stwob\ to
less than 0.1 and allowing searches for other manifestations of
\CP\ violation in the $B$ system.


\end{document}